\documentclass[a4paper,11pt]{article}
\pdfoutput=1 % if your are submitting a pdflatex (i.e. if you have
\usepackage{jinstpub} % for details on the use of the package, please
\usepackage{ads}
\usepackage{natbib}

\usepackage{lineno}
\usepackage{siunitx}
\usepackage[nooneline]{subfigure}
\subfiguretopcaptrue

\usepackage{enumerate}

\usepackage{xcolor}

\usepackage{ulem}
\DeclareRobustCommand{\erase}{\bgroup\markoverwith{\textcolor{orange}{\rule[.5ex]{1.5pt}{1.5pt}}}\ULon}

\title{\boldmath High-speed Readout System of X-ray CMOS Image Sensor for Time Domain Astronomy}

\author[a,1]{Naoki Ogino\note{Corresponding author.}}
\author[a,2]{Makoto Arimoto\note{Corresponding author.}}
\author[a]{Tatsuya Sawano}
\author[a]{Daisuke Yonetoku}
\author[c]{Hsien-chieh Shen}
\author[c]{Takanori Sakamoto}
\author[b]{Junko S. Hiraga}
\author[d]{Yoichi Yatsu}
\author[e]{Tatehiro Mihara}

\affiliation[a]{Kanazawa University, Kakuma, Kanazawa, Ishikawa, Japan}
\affiliation[b]{Aoyama Gakuin University, 5-10-1 Fuchinobe, Chuo-ku, Sagamihara, Kanagawa, Japan}
\affiliation[c]{Kwansei Gakuin University, 2-2 Gakuen, Sanda, Hyogo, Japan}
\affiliation[d]{Tokyo Institute of Technology, 2-12-1 Ookayama, Meguro-ku, Tokyo, Japan}
\affiliation[e]{Institute of Physical and Chemical Research (RIKEN), 2-1, Hirosawa, Wako, Saitama 351-0198, Japan}

\emailAdd{naokiogino@stu.kanazawa-u.ac.jp}
\emailAdd{arimoto@se.kanazawa-u.ac.jp}

\abstract{
We developed an FPGA-based high-speed readout system for a complementary metal-oxide-semiconductor (CMOS) image sensor to observe soft X-ray transients in future satellite missions, such as \textit{HiZ-GUNDAM}. 
Our previous research revealed that the CMOS image sensor has low-energy X-ray detection capability (0.4–4\,keV) and strong radiation tolerance, which satisfies the requirements of the \textit{HiZ-GUNDAM} mission. 
However, CMOS sensors typically have small pixel sizes (e.g., $\sim$10\,\si{\micro m}), resulting in large volumes of image data. 
GSENSE400BSI has 2048$\times$2048 pixels, producing 6\,Mbyte per frame. 
These large volumes of observed raw image data cannot be stored in a satellite bus system with a limited storage size. 
Therefore, only X-ray photon events must be extracted from the raw image data. 
Furthermore, the readout time of CMOS image sensors is approximately ten times faster than that of typical X-ray CCDs, requiring faster event extraction on a timescale of $\sim$0.1\,s. 
To address these issues, we have developed an FPGA-based image signal processing system capable of high-speed X-ray event extraction onboard without storing raw image data.
The developed compact system enabled mounting on a CubeSat mission, facilitating early in-orbit operation demonstration. 
Here, we present the design and results of the performance evaluation tests of the proposed FPGA-based readout system. 
 Utilizing X-ray irradiation experiments, the results of the X-ray event extraction with the onboard and offline processing methods were consistent, validating the functionality of the proposed system.
}


\proceeding{24$^{\text{th}}$ International Workshop on Radiation Imaging Detectors\\
  25-29 June, 2023\\
  Oslo, Norway}

\begin{document}
\maketitle
\flushbottom

\section{Introduction}
\label{sec:intro}
A future satellite mission, \textit{HiZ-GUNDAM} \cite{Yonetoku+2020}, is being developed for time-domain astronomy, covering an energy band range of 0.4–4\,keV. 
This mission has multiple payloads, including a wide-field X-ray monitor to detect X-ray transients such as gamma-ray bursts \cite{Klebesadel+1973} and an infrared telescope specifically designed to perform follow-up observations of transients detected by the wide-field X-ray monitor \cite{2020SPIE11443E..0RT}.
Here, a wide-field X-ray monitor was designed with a field of view of ${\sim}1$\,str and sensitivity of $\sim$$10^{-10}$\,\si{erg.s^{-1}. cm^{-2}} for a 100-s exposure. 
The monitor also localizes the direction of the X-ray transients with a spatial resolution of $\sim$2\,arcmin. 
Considering that the typical duration of gamma-ray bursts is $\sim$0.1–100 s \cite{Meegan+1996}, a photon-counting capability with a time resolution of $\leq$0.1\,s is required.

The wide-field X-ray monitor was equipped with lobster eye optics \cite{Angel+1979, 2022SPIE12181E..5JG} to satisfy the wide field of view and high-sensitivity requirements and a focal pixel detector that can acquire the position and energy of incident X-ray photons.
A back-illuminated complementary metal-oxide-semiconductor (CMOS) sensor \cite{Fossum+1997} was positioned on the focal plane as an X-ray pixel detector.
Our previous study confirmed that the CMOS sensor GSENSE400BSI-TVISB (Gpixel Inc., Table \ref{tab:gsense400bsi}) \cite{Ma+2015,Wang+2018,Harada+2019,Narukage+2020} is well-suited for mission requirements because of its low-energy X-ray detection capability, fast readout speed, and radiation tolerance \cite{Ogino+2021}.

Owing to the limitations of data storage onboard satellites, storing all the observed image data from CMOS sensors with an extremely large number of pixels (e.g., 2000 $\times$ 2000 pixels produces 6-Mbyte per frame for 12 bits) is impossible.
Because the ratio of pixels with X-ray photon events to all pixels is less than 0.01\%, a significant reduction in data size must be achieved by selectively extracting only X-ray events from the raw image data. 
Furthermore, this electronic readout system should be compact because of the limited space available on the satellite.
Therefore, we developed a compact readout system dedicated to a CMOS sensor that performs X-ray event extraction and is suitable for satellite applications.

\begin{table}
  \centering
  \caption{Specifications of GSENSE400BSI-TVISB}
  \begin{tabular}{ll} \hline
     Items             & Parameters                          \\ \hline \hline
     Active image size & $22.528{\times}22.528$\,$\mathrm{mm^2}$ \\
     Pixel size        & $11{\times}11$\,\si{\micro m^2}     \\
     Number of pixels  & $2048{\times}2048$                  \\
     Max frame rate    & 48\,fps                             \\
     Pixel clock rate  & 25\,MHz                             \\
     Power consumption & $<$0.650\,W                         \\ \hline
  \end{tabular}
  \label{tab:gsense400bsi}
\end{table}

\section{Hardware Design and Functions}
\begin{figure}
  \centering
  \includegraphics[width=0.5\textwidth]{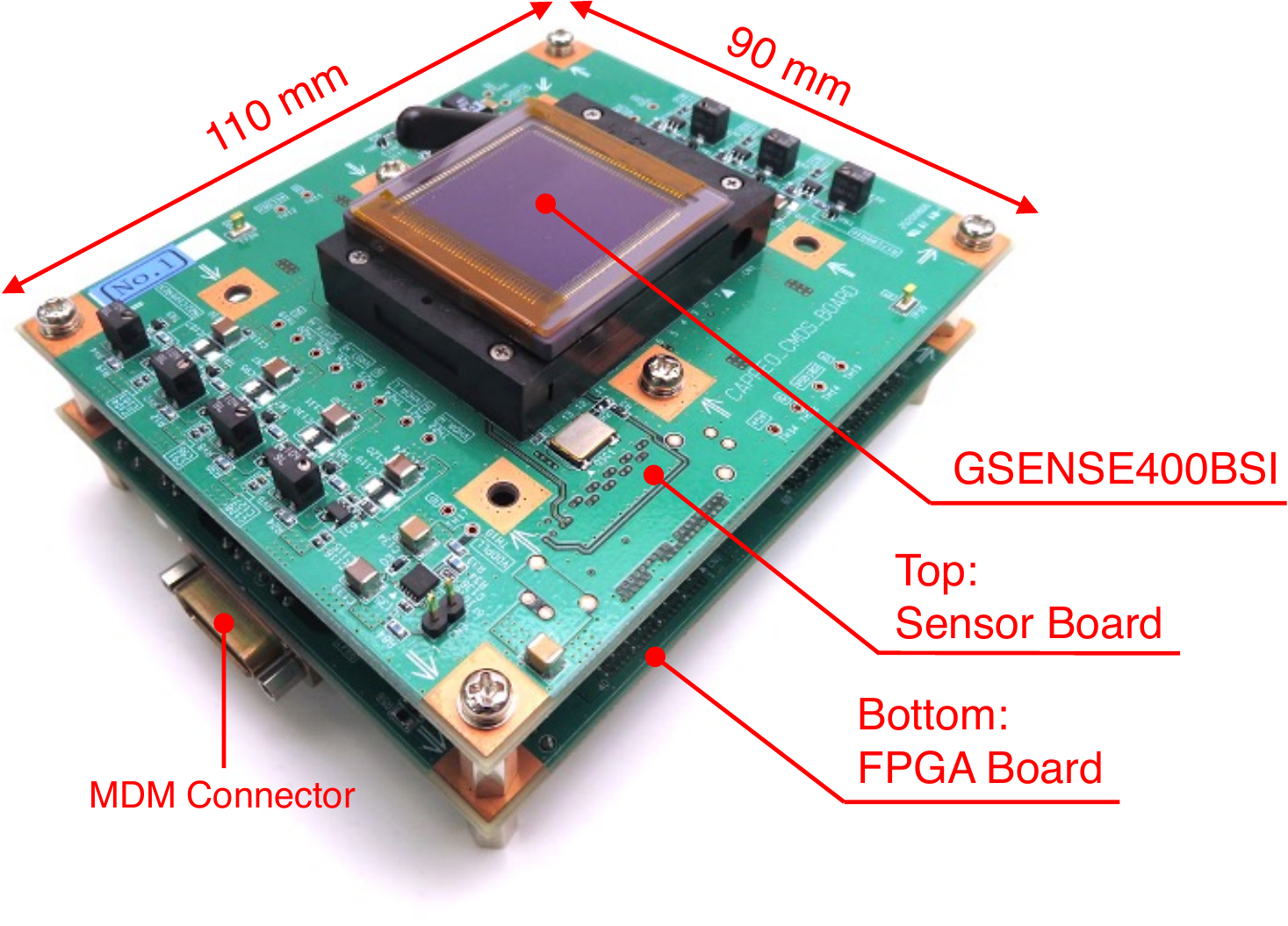}
  \caption{Picture of our readout system, which comprises two boards: FPGA board (bottom) and CMOS sensor board (top). CMOS Board is 90\,mm$\times$110\,mm$\times$39\,mm with CMOS sensor mounted and can fit into dimensions of 2U CubeSat. FPGA board has MDM connector to supply power and input/output commands and data packets.}
  \label{fig:capreo}
\end{figure}

% 基板サイズや使用したFPGAについて
We adopted image processing utilizing a field-programmable gate array (FPGA) \cite{Brown+1996} to achieve high-speed synchronous image acquisition control from the CMOS sensors. The developed readout system comprises an FPGA board and a CMOS sensor board with stack structure and a compact size of 90\,mm $\times$ 110\,mm $\times$ 39\,mm combined with the CMOS sensor (Figure~\ref{fig:capreo}). This system
can be installed on middle-class satellites (e.g., \textit{HiZ-GUNDAM}) and nanosatellites, such as the 6U CubeSat developed by the Tokyo Institute of Technology \cite{Yatsu+2018}.
The FPGA board was equipped with a Xilinx Kintex-7 (XC7K325T-1FFG676I) with 400 I/Os \cite{Kintex-7_IOs} and an internal block RAM of $\sim$2MB \cite{Kintex-7}, providing sufficient capabilities for controlling the CMOS sensor, reading out image data from the CMOS sensor, and extracting X-ray events from the raw image data, as mentioned in Sec. \ref{sec:event_extraction}. 
This system operates with an external power supply of +5\,V and outputs of +1.8\,V, +3.3\,V etc., utilized for digital ICs, including analog-to-digital converters (AD7928BRUZ, Analog Devices) and the FPGA.

% モードの定義
We implemented the following three modes to be common for various observations with the CMOS sensor.
\begin{enumerate}[(1)]
\item Frame mode: This mode outputs all the raw image data and is mainly employed for the optical and ultraviolet observations.
\item Event mode: This mode extracts only X-ray photon events from the observed image data and outputs their position, energy and time information, mainly utilized for X-ray photon-counting observations.
\item Housekeeping mode: In this mode, the housekeeping data of this system (e.g., the temperatures of the CMOS sensor and the FPGA board, the values of the voltage and current supplied to the CMOS and FPGA boards) are integrated as data packets and transferred to the bus system. This mode is adopted to monitor the system's health status.
\end{enumerate}

\begin{figure}
  \centering
\includegraphics[width=0.82\textwidth]{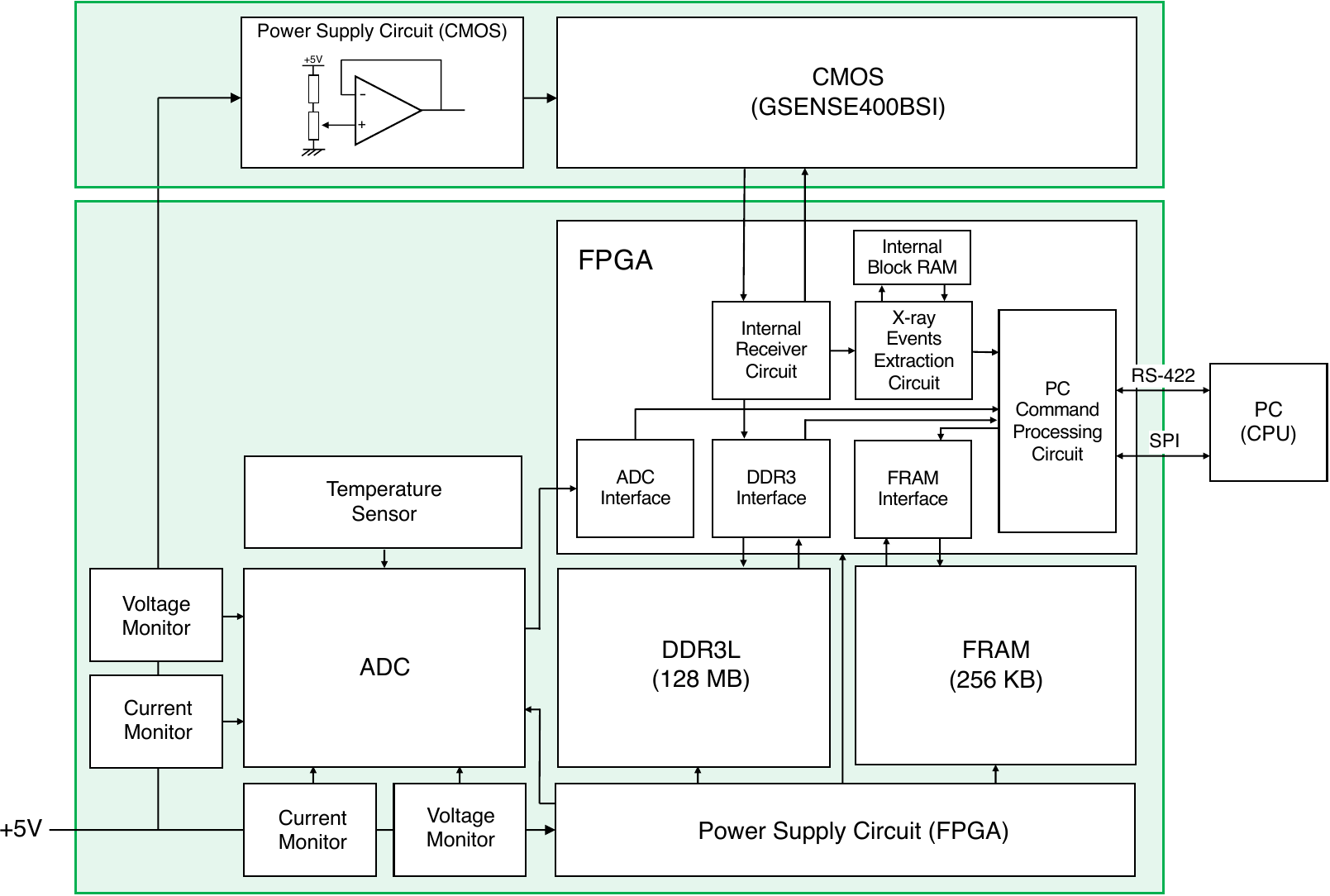}
  \caption{Block diagram of readout system configuration. System operates at an external power supply of +5\,V. Temperatures of CMOS sensor and FPGA board, measured voltages and currents of two boards, are constantly monitored to check system's health status. Extraction of X-ray events is performed with internal block RAM; DDR3L is utilized as buffer for output of frame image data. RS-422 and SPI communication are adopted for FPGA and external personal computer (PC) communication.}
  \label{fig:system-config}
\end{figure}

% DDR3の用途
The system configuration of the developed readout board is illustrated in Figure \ref{fig:system-config}. The FPGA board has two types of memory:128\,MB DDR3L synchronous dynamic random access memory (SDRAM) \cite{Nikel+1999} for storing raw image data from the CMOS sensor and 256\,KB ferroelectric random access memory (FRAM) \cite{Ishiwara+2012} for storing the configuration data of the CMOS sensor. The DDR3 memory functions as a buffer during operation in frame mode. Here, the power supply to the DDR3 memory can be turned off to reduce power consumption in event mode.
% FRAM
FRAM, a type of nonvolatile memory that stores data utilizing ferroelectric polarization, is resistant to bit flipping caused by cosmic rays, known as single-event upset (SEU) \cite{Nicolaidis+2005}. Therefore, it stores critical operational settings for satellite missions, such as the CMOS setting data and threshold values required for extracting X-ray events that are not frequently rewritten. In orbit, for protection from bit flipping by the SEU, the setting data read from the FRAM are stored in three areas of the FPGA, and the majority decision is then applied.

% 通信
This system incorporates two communication lines: the control line, which is utilized to send and receive commands such as changes in the operating mode, updating the configuration values, and sending back acknowledgments. Because high-speed data communication was not necessary for this line, RS-422 was adopted. The second is a data line, which transmits the frame and event data. Because the data line requires faster communication than the control line, an SPI communication of a few megabytes per second was utilized.

\section{X-ray Event Extraction utilizing FPGA}\label{sec:event_extraction}
% イベント抽出
This section describes the procedure for the X-ray event extraction implemented in the FPGA.
Upon the interaction of X-rays with the CMOS sensor, X-ray events can be divided into two pixel patterns: "single-pixel" and "multi-pixel" \cite{Ogino+2021}, because signals on the pixels from an X-ray photon are often distributed within a single pixel and multi-pixels. 
In the case of multi-pixel events, the correct energy information for an incident X-ray photon can be obtained by summing the pixel values.

\begin{figure}
  \centering
\includegraphics[width=0.72\textwidth]{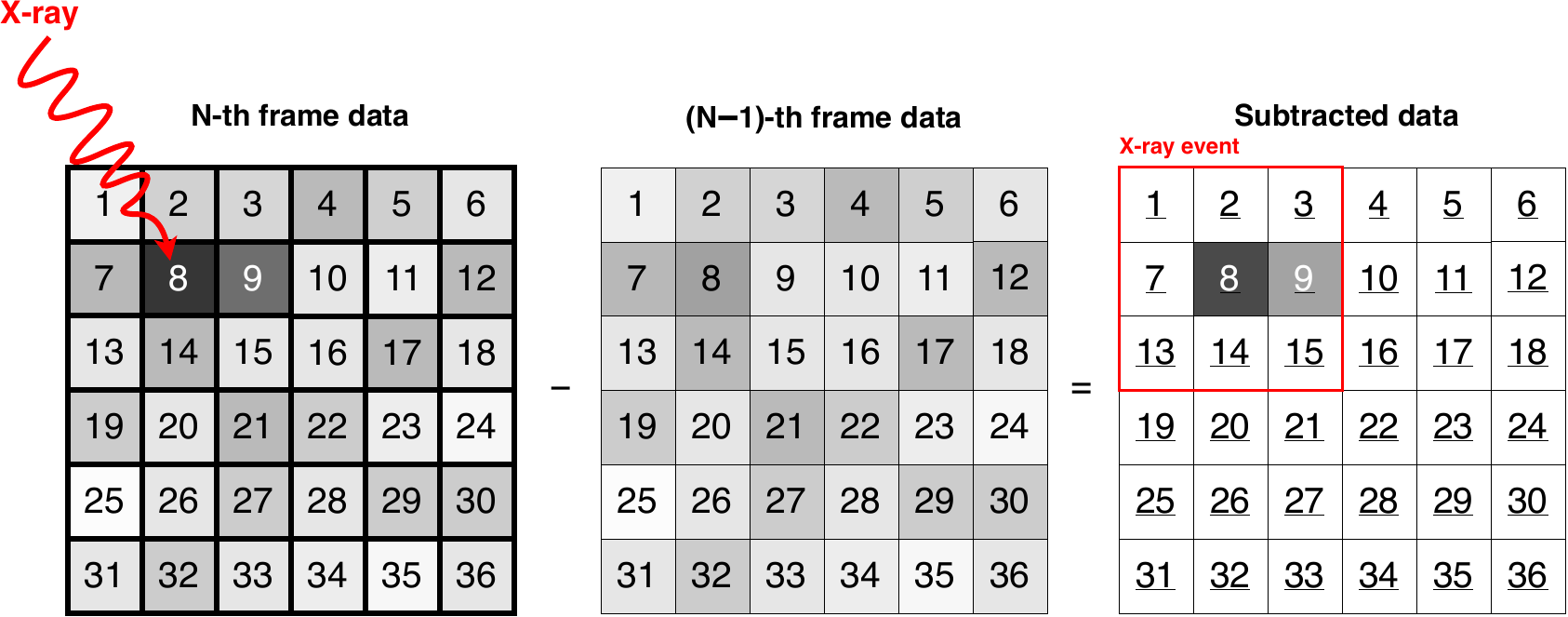}
  \caption{Schematic of X-ray event extraction with offline procedure:
            The number for each pixel represents the pixel ID. The ADU values, represented in grayscale, stored in the frame buffer differ for each pixel and vary with the operation temperature and radiation damage. To remove this effect and identify X-ray events,
           we take the difference between the $N$-th image data and the $(N-1)$-th image data. Then, X-ray photon events can be efficiently obtained without storing all of the raw image data. The underlined numbers represent the pixel ID with subtracted data. }
  \label{fig:offline}
\end{figure}

% オフラインソフトでの抽出方法
We first describe the event extraction procedure performed utilizing offline software.
For simplicity, we consider a CMOS sensor with a 6$\times$6 pixel area.
To process the image data and identify X-ray events with photon energies and positions, we adopted the following method in the offline software (Figure \ref{fig:offline}):
\begin{enumerate}[(1)]
    \item In the image data output from the CMOS sensor, each pixel has an offset voltage. This voltage varies for each pixel owing to the CMOS fabrication processes. Thus, for obtaining X-ray photon energy accurately and efficiently, it is essential to subtract the consecutive pixel images, e.g., subtraction between the 2nd and 1st images, 3rd and 2nd images, 4th and 3rd images, and so forth.
    \item We adopted two types of threshold to detect only the X-ray photon events efficiently: an "event threshold" to identify a primary pixel which mainly interacted with an X-ray photon and a "split threshold" to collect signals from the adjacent 3$\times$3 pixels around the primary pixel for the multi-pixel events.
    \item Pixels with analog-to-digital-unit (ADU) values above the event threshold are identified as X-ray events.
    \item If the signals from any adjacent pixel do not exceed the split threshold, the event is labeled as a single-pixel event. Otherwise, the event is categorized as a multi-pixel event, and all ADU values exceeding the two thresholds in the 3$\times$3 pixels are summed up to derive the X-ray photon energy.
\end{enumerate}

\begin{figure}
  \centering
  \includegraphics[width=0.60\textwidth]{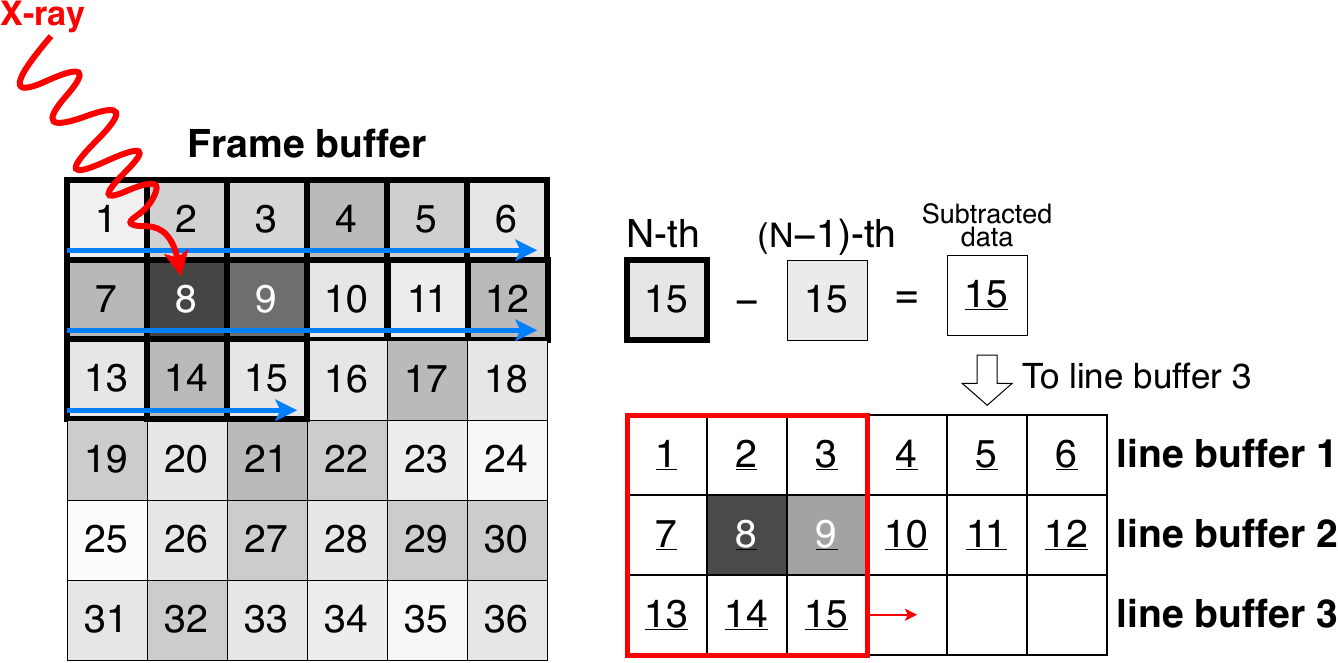}
  \caption{Schematic of X-ray event extraction with onboard procedure:
             Number of each pixel and grayscale are same as in Figure~\ref{fig:offline}. Subtracted data are allocated to three line buffers in order. When ADU values of 3$\times$3 pixels are obtained, system performs pixel pattern matching developed by \textit{ASCA}/SIS method to derive plausible X-ray photon events.}
  \label{fig:buffer}
\end{figure}

% オンボードでの抽出方法
%\erase{To realize the above processing onboard, one frame buffer and three line buffers were implemented in the internal block RAM of the FPGA.} 
The following procedure was implemented on the FPGA. The schematic procedure of the FPGA-based real-time event processing is illustrated in Figure~\ref{fig:buffer}.
Here, owing to the limitation of the size of the internal block RAM of the FPGA, two-frame buffers for storing two raw image datasets cannot be implemented. Instead, we utilized a one-frame buffer to store two raw image data points and three line buffers with a small RAM size for subtracting two images (see below for details). 
The numbers in the frame buffer in Figure~\ref{fig:buffer} depict the pixel IDs, and the sensor outputs data in order from the 1st to the 36th pixel.

\begin{enumerate}[(1)]
    \item Figure~\ref{fig:buffer} presents just one example with subtraction of pixel ID 15 to illustrate the procedure. First, the $N$-th image data is stored in pixel IDs 1–14, and the $(N-1)$-th image data is stored in pixel IDs 15–36, as illustrated in Figure~\ref{fig:buffer}. When the ADU value with pixel ID 15 is taken from the sensor, the system subtracts the $(N-1)$-th data stored in the frame buffer from the ADU value taken from the sensor with pixel ID 15, and the subtracted ADU value is obtained.
    \item After subtracting pixel ID 15, the ADU value of pixel 15 in the frame buffer is updated to the ADU value taken from the sensor.
    \item The subtracted data is stored in the line buffer. Then, the system performs pixel pattern matching with 3$\times$3 pixel data for X-ray event extraction. Here, we follow the pattern matching developed by \textit{ASCA}/SIS of the previous astronomical satellite \cite{Burke+1994}.
    \item By repeating steps (1)–(3), event extraction of all image data can be performed. Here, if the three line buffers are filled at pixel ID 18, a new pixel value of pixel ID 19 is stored at the starting point of the line buffer 1. The same procedure is performed subsequently but with a modified pixel pattern due to a different row position.
\end{enumerate}
In the next section, we present the results of system performance verification employing the two methods. 

\section{Performance verification}
% 性能評価
To confirm that the system with the implemented functions functioned properly, the following two experiments were performed: 
\begin{enumerate}[(1)]
  \item Real-time detection and event extraction with an exposure time of 0.1\,s by mimicking a transient X-ray phenomenon utilizing a radioactive source. 
  \item Comparison of spectra obtained by the offline and onboard processing methods and consistency verification between them. 
\end{enumerate}

\begin{figure}
  \centering
  \subfigure[Setup]{
\includegraphics[width=0.40\textwidth]{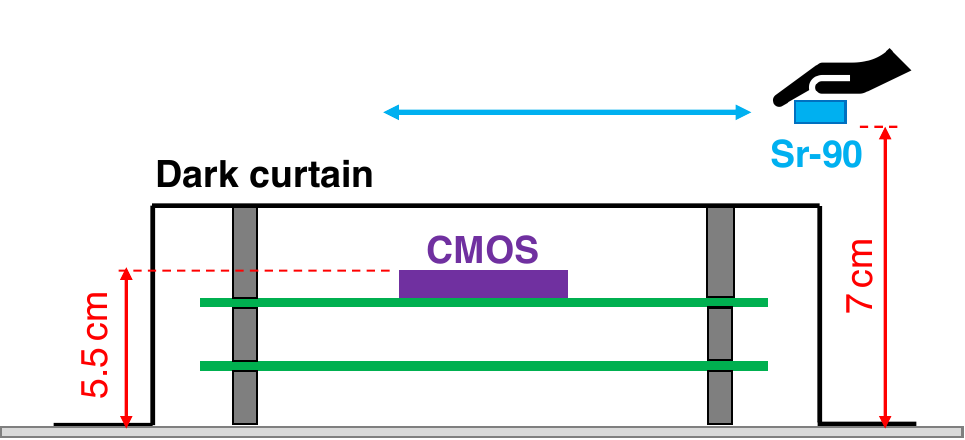}
      \label{fig:capreo_Sr}
  }
  \subfigure[Light curve]{
\includegraphics[width=0.45\textwidth]{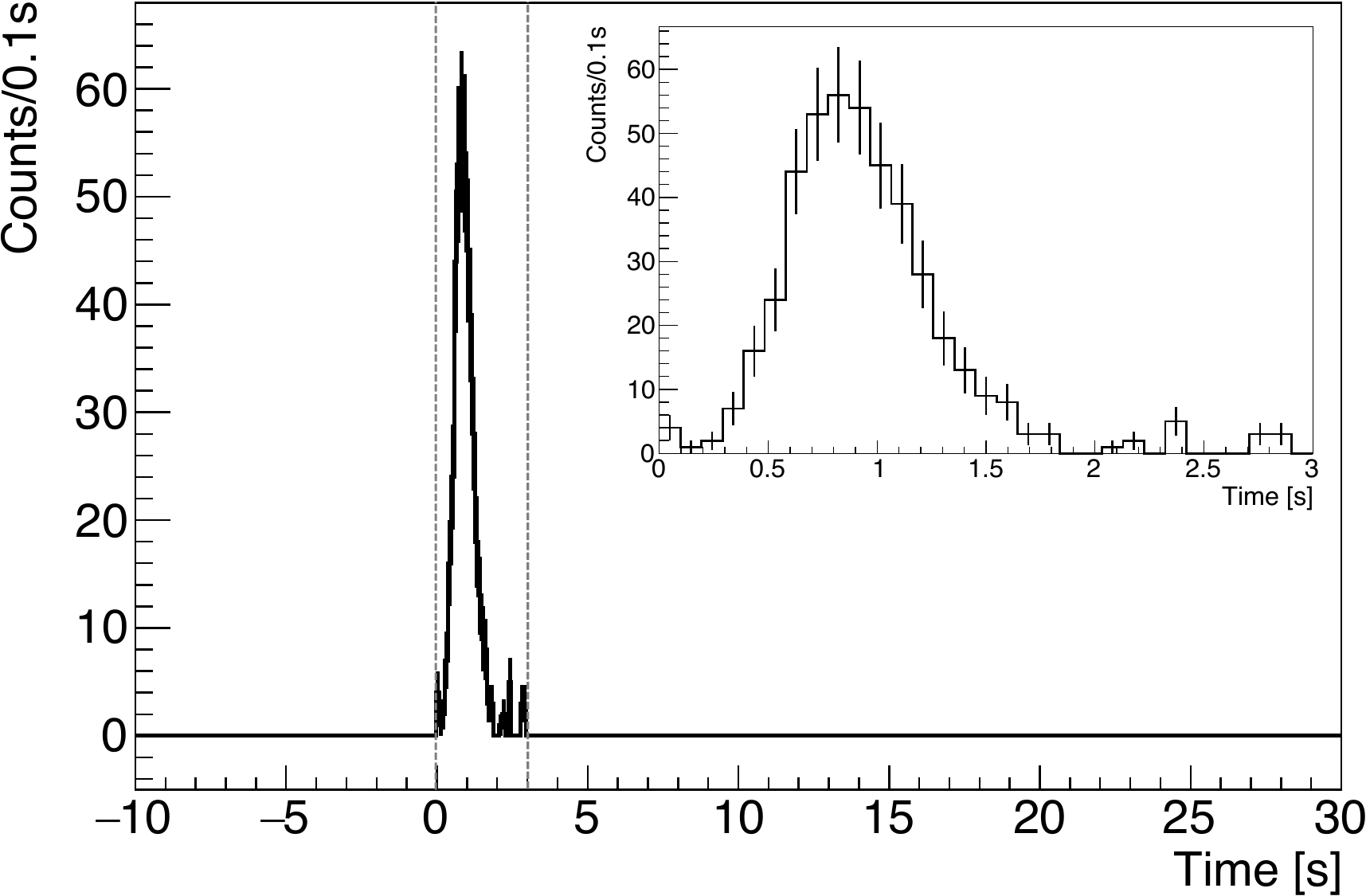}
      \label{fig:capreo_lc}
  }
  \caption{(a) Schematic of experimental setup for obtaining light curves. Light curve was acquired by covering developed readout system and CMOS sensor with dark curtain and quickly moving Sr-90 over it at room temperature.
           (b) Acquired light curves with count of $\beta$-rays emitted from Sr-90 at exposure time of 0.1\,s. Inset figure is zoomed view from 0\,s to 3\,s (dashed line interval).}
\end{figure}

% ライトカーブの取得
For the first experiment, we utilized $\beta$-rays emitted from Sr-90 with $\sim$$1\times10^4$\,Bq in the air to mimic an X-ray transient event with a timescale of $\sim$1 s. Because of low radioactivity, mimicking a clear X-ray pulse utilizing X-ray radioactive sources was difficult.
Sr-90 beta decays to Y-90 with a half-life of 28.79\,years, emitting a maximum of 546 keV $\beta$-rays. Y-90 beta decays to Zr-90 with a half-life of 64\, h, emitting a maximum of 2.28\,MeV $\beta$-rays \cite{HANSEN+1983}. 
$\beta$-rays emitted by these decay processes enter the CMOS sensor and deposit energies of $\sim$4 keV. The deposited energy is almost equivalent to the target X-ray energy band of \textit{HiZ-GUNDAM}. Note that $\sim$99\% of $\beta$-ray events are multi-pixel events.
Figure~\ref{fig:capreo_Sr} illustrates the experimental setup. The CMOS sensor and the developed readout system were covered with a dark curtain. 
Subsequently, a light curve measured every 0.1\,s was obtained by moving Sr-90 quickly over the CMOS sensor. Figure~\ref{fig:capreo_lc} presents the measured light curve obtained utilizing the onboard detection method. This result indicates that a clear spike was detected with a timescale of $<$1 s and that the FPGA-based event processing worked correctly.

\begin{figure}
  \centering
  \includegraphics[width=0.48\textwidth]{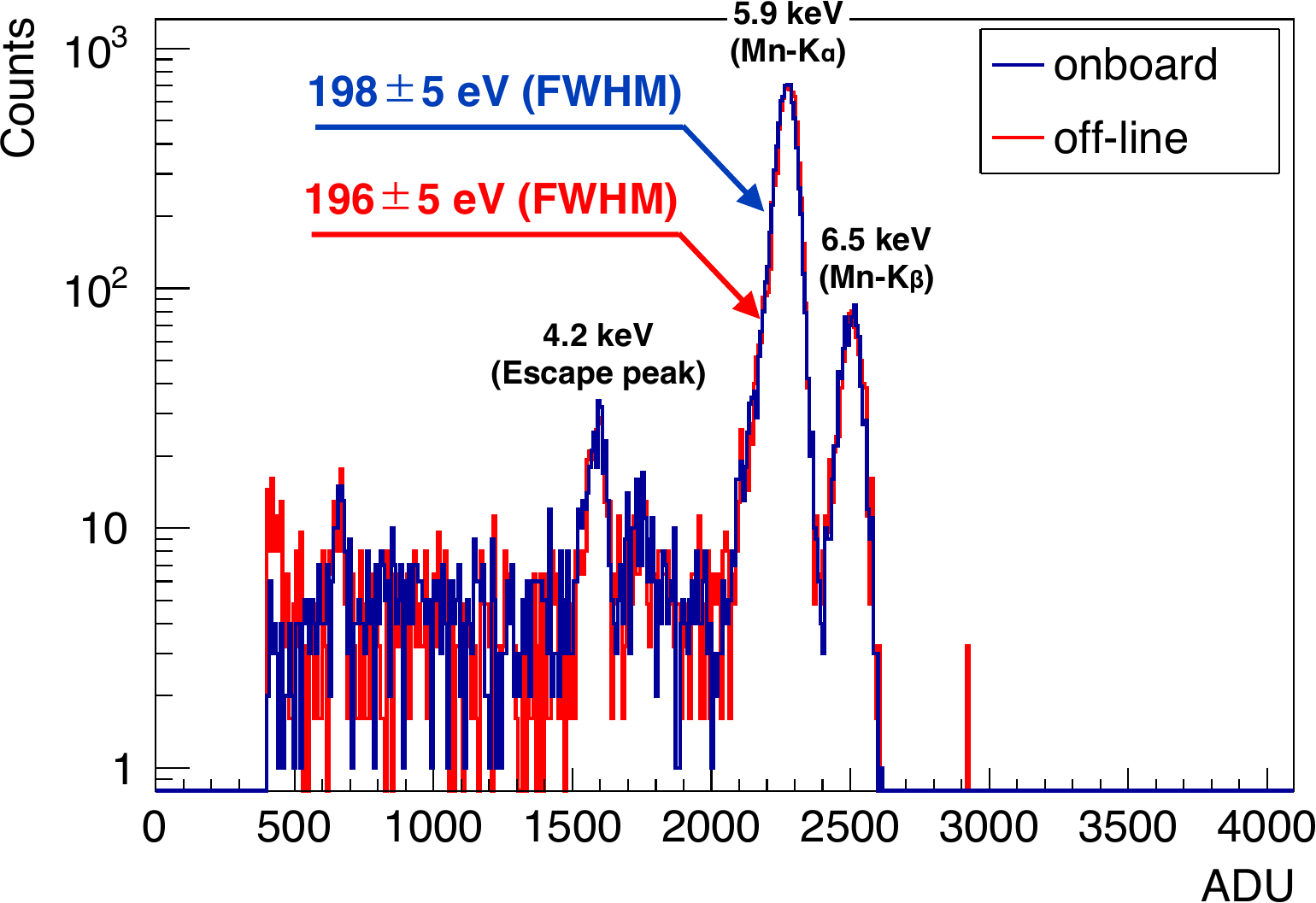}
  \caption{Spectra of ${}^{55}\mathrm{Fe}$ source (5.9\,keV and 6.5\,keV) at $+$20$^\circ$C.
  Blue and red lines denote obtained spectra by onboard processing with this system and offline analysis with full-frame images, respectively.}
  \label{fig:spectrum}
\end{figure}

% スペクトルの取得
In the second experiment, we measured the 5.9\,keV (Mn-K$_{\alpha}$) and 6.5\,keV (Mn-K$_{\beta}$) lines from Fe-55 adopting this system.
Figure~\ref{fig:spectrum} illustrates the spectra obtained utilizing offline and onboard methods for an Fe-55 source at $+$20\,$^\circ$C.
The X-ray fluorescence lines at 5.9\,keV and 6.5\,keV from Fe-55 and 
Si escape line of Mn-K$_{\alpha}$ (4.2\,keV) were clearly detected.
Furthermore, the two measured spectra obtained via the two methods were identical within statistical uncertainty.
The measured energy resolutions at 5.9\,keV (full width at half maximum, FWHM) were 196$\pm$5\,eV utilizing the offline method, and 198$\pm$5\,eV the onboard method. 
This result confirms that there is no significant noise contribution due to the different processing methods and that the onboard acquisition system can obtain accurate X-ray energy information without increasing noise.

Note that the power consumption of the FPGA for the two experiments was measured to be approximately 3\,W, except for the CMOS sensor.

\section{Conclusion}
We developed a CMOS image-sensor readout system for future satellite missions \textit{HiZ-GUNDAM}. This system has a stacking structure comprising a CMOS sensor and FPGA boards with a compact size that can be mounted on the 2U-CubeSat. Adopting an FPGA-based X-ray event-extraction system, the light curve and spectrum were successfully measured in real time without significant performance degradation. % consistent with the offline method adopting the full-frame data %acquired by utilizing the radiation emitted from the radiation source to the completed system, and it was
Our results demonstrate that the developed high-speed readout system works well and applies to future satellite missions.

\acknowledgments
This study was supported by JSPS KAKENHI Grant Numbers JP17H06362 (M.A.), JP22J12717 (N.O.), JP23H04898 (M.A. and D.Y.), JP23H04895 (T. Sawano and T. Sakamoto), the CHOZEN Project of Kanazawa University (M.A., T. Sawano, and D.Y.), and the JSPS Leading Initiative for Excellent Young Researchers Program (M.A.).

\bibliographystyle{unsrt}
\bibliography{main}

\end{document}